# ChatGPT as a tool for honing teachers' Socratic dialogue skills

Bor Gregorcic[1], Giulia Polverini[1], Andreja Sarlah[2]

[1] Department of Physics and Astronomy, Uppsala University, Box 516, 75120 Uppsala, Sweden
[2] Faculty of Mathematics and Physics, University of Ljubljana, Jadranska 19, 1000 Ljubljana, Slovenia

E-mail: bor.gregorcic@physics.uu.se

**Abstract**
In this proof-of-concept paper, we propose a specific kind of pedagogical use of ChatGPT – to help teachers practice their Socratic dialogue skills. We follow up on the previously published paper "ChatGPT and the frustrated Socrates" by re-examining ChatGPT's ability to engage in Socratic dialogue in the role of a physics student. While in late 2022 its ability to engage in such dialogue was poor, we see significant advancements in the chatbot's ability to respond to leading questions asked by a human teacher. We suggest that ChatGPT now has the potential to be used in teacher training to help pre- or in-service physics teachers hone their Socratic dialogue skills. In the paper and its supplemental material, we provide illustrative examples of Socratic dialogues with ChatGPT and present a report on a pilot activity involving pre-service physics and mathematics teachers conversing with it in a Socratic fashion.

Keywords: *chatbots, ChatGPT, GPT-4, GPT-3.5, Socratic dialogue, teacher training, pedagogical skills*.

**1 Introduction**

Only a year after the release of ChatGPT, the use of Artificial Intelligence-based chatbots in education has gained steam. This upward trend is expected to continue due to the constantly growing body of research on the potential of this technology to improve teaching and learning [1–3]. Despite the risks related to the use of AI-based chatbots [4,5], they hold the potential to be useful for both students and teachers. Students could benefit from having a personal, always-available, and patient tutor [6,7], a peer for doing homework [8], or an "object to think with" [9]. On the other hand, teachers could take advantage of a time-saving assistant for tasks such as grading [10], or lesson planning [11], as well as a source of material for classroom or homework activities [12].
However, for these applications to achieve optimal effectiveness, it is crucial that the chatbot performs well in the specific domain [13]. This does not seem to be the case in physics, where even the state-of-the-art chatbot ChatGPT-4[1] [14] still has some way to go before its performance can be considered expert-like. Surprisingly, this is true not only for advanced physics topics but even for introductory conceptual physics, where it can still clumsily fail [15]. Therefore, the effectiveness of the technology is contingent upon the users' expertise in the subject matter, their ability to craft useful prompts, and critically evaluate the outputs. In this paper, we explore a potential application of ChatGPT for training physics teachers in a specific pedagogic skill. To become experts at facilitating learning, teachers need to have the ability to engage in meaningful dialogue with students. Above all, they need to practice "active listening", recognising the strengths and weaknesses of students' reasoning and responding in appropriate ways [16]. One technique by which a teacher can help students is not directly validating or invalidating their answers, but instead asking strategically placed questions, which can lead students to improve their answers. We refer to this approach as "Socratic dialogue". Here we discuss the feasibility of several versions of ChatGPT to play the role of a tutee, or a *model of a student*, with whom teachers can engage in Socratic dialogue. This may be potentially useful for pre-service teachers, who often do not have easy access to students with whom they could practice their teaching skills. Moreover, because of the lack of experience, it is easier for pre-service teachers to get overwhelmed by the complexity of real-world teaching situations. The controlled, private and safe environment offered by a chatbot could potentially serve as a good starting point.

## 2 Background

In late 2022, just a few weeks after the release of ChatGPT, one of the authors, together with a colleague, reported in this journal on their initial experimentation with this chatbot, asking it a conceptual introductory physics question:

> *"A teddy bear is thrown into the air. What is its acceleration in the highest point?"* [17] (p. 2).

They noticed that the response was linguistically advanced, but contained serious physics errors and inconsistencies. Moreover, upon attempting to have the chatbot reconcile the inconsistencies through Socratic dialogue, they were met with confident-sounding but incorrect responses, a feature well-recognized in the model that was in use at the time [18]. They also found it very difficult to have the chatbot notice its own errors and develop its responses based on a critical reassessment of its own answers, even when they were clearly incorrect and inconsistent with each other. The first conclusion at the time was that the fairly common generation of physics errors and inconsistencies made ChatGPT unsuitable for the role of a physics tutor. Furthermore, its confidence and stubbornness severely limited the possibility of having a Socratic dialogue, as it happens between a teacher and a student. Thus, the title of the paper "ChatGPT and the frustrated Socrates".

Since then, ChatGPT has evolved in terms of both its performance in physics, and its "character" during dialogue. Both the freely available version (ChatGPT-3.5) and the subscription-based one (ChatGPT-4) have experienced improvement. For example, ChatGPT-3.5 no longer makes the same mistake in answering the above-cited question, and has become "humbler" and more responsive to the user's follow-up questions. For an illustration of a Socratic-style dialogue with ChatGPT-3.5 done in February 2024, see supplemental material A1. On the other hand, ChatGPT-4 outperforms ChatGPT-3.5 in terms of its physics performance [19,20] and, as we show in this paper, its responsiveness to critical feedback and guiding questions. We illustrate this in section 3, through a case study of a Socratic interaction with ChatGPT-4. For another example of a Socratic dialogue with ChatGPT-4, see supplemental material A2.

---
[1] ChatGPT-4 is a shorthand for ChatGPT powered by the GPT-4 large language model, the latest version released by OpenAI in March 2023.

The finding that both versions have become better partners in Socratic dialogue allows us to revisit the idea of using ChatGPT as a training tool for pre-service teachers. In section 4, we report on a practical implementation of a Socratic dialogue-based activity with ChatGPT-3.5 in the context of pre-service physics teacher training, carried out in April 2023. We also highlight the immediate benefits and challenges of using the chatbot in this way. Examples of pre-service teachers' conversations with ChatGPT can be found in supplemental material B.

In section 5, we address the potential of the two ChatGPT's versions to serve as a *model of a student* for the purposes of teacher training, and reflect on the experienced and potential benefits of such activities.

Finally, in section 6 we summarise the key findings and provide suggestions for future work on the topic.

**3 Socratic dialogue with ChatGPT-4: an illustrative example**

To be able to engage in Socratic dialogue, we first looked for tasks that the chatbot would not solve correctly on the first attempt[2]. For example, we found that the question used in the "Frustrated Socrates" was too easy for ChatGPT-4: we asked it 10 times in separate chat windows and always received a correct answer. We thus used a slightly more challenging but topically related question, which turned out to be difficult enough for the chatbot to answer incorrectly:

> *Nina releases a bouncy ball, which falls to the ground and bounces back to the same height it was released from. Describe the graph of velocity vs. time for the ball.*

Figure 1 shows our own sketch of the v-t graph for the described scenario. We have chosen the positive direction to be upwards. We see that the velocity starts at zero, as the ball is released from rest. The ball first accelerates downwards with constant acceleration (directed downward and thus negative). This is represented by the straight line with constant negative slope, starting at $t=0$ and reaching its minimum (maximum in speed) just before the collision with the ground ($t=t_c$). We assume that the collision happens instantaneously and is perfectly elastic. At the moment of collision, the velocity changes direction and becomes positive, with the same magnitude it had just before the collision (assuming a perfectly elastic collision). At the moment of collision ($t=t_c$), the graph thus "jumps" from the most negative velocity to the most positive velocity (of the same magnitude). Just after the collision, the ball moves upward (in the positive direction), having positive velocity, which decreases linearly at the same rate as before the bounce (with the same negative constant acceleration). Our observation ends when the ball's velocity becomes zero, indicating it has momentarily stopped at the highest point (at the height of its initial release). After this, the straight downward-sloping line of the graph would continue until the next bounce.

---

[2] Although this is not strictly necessary. We can engage in a Socratic dialogue with the chatbot even when it answers correctly, in order to probe it further.

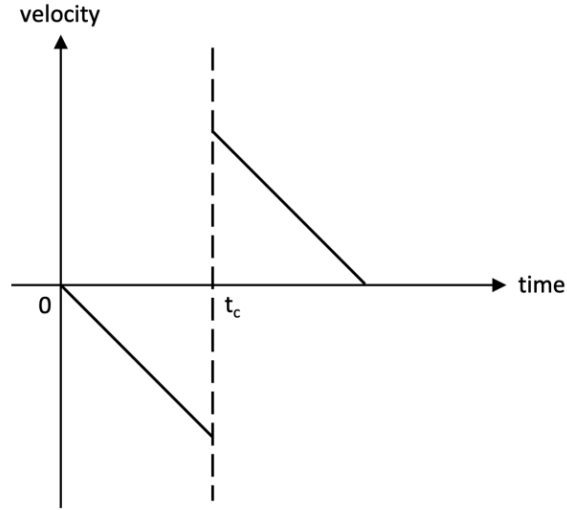

**Fig.1:** A sketch of the velocity vs. time graph of a ball falling from rest and bouncing from the floor to the same height it was released. The dashed vertical line at $t_c$ marks the moment of collision with the ground and highlights the discontinuity of the function.

We now turn to ChatGPT-4's initial response to the above-stated task. The excerpt below shows our prompt, followed by the chatbot's response. We bolded the incorrect statements.

> **User**
>
> Nina releases a bouncy ball, which falls to the ground and bounces back to the same height it was released from. Describe the graph of velocity vs. time for the ball.
>
> **ChatGPT**
>
> The graph of velocity vs. time for a bouncy ball that falls to the ground and bounces back to the same height would have the following characteristics:
>
> 1. **Initial Descent Phase**: At the start, when Nina releases the ball, it starts from rest, so its initial velocity is zero. As it falls, its velocity increases in the negative direction (assuming down is negative) due to gravity. This section of the graph would show a straight line with a negative slope, representing constant acceleration (due to gravity).
>
> 2. **Just Before Impact**: Right before the ball hits the ground, its velocity is at its maximum (in the negative direction), as it has been accelerating downwards.
>
> 3. **Impact with the Ground**: The impact with the ground is a very brief moment where the velocity changes direction. The graph here would show a sharp spike, as the velocity changes from its maximum negative value to a positive value. This is an instantaneous change, representing the elastic collision with the ground.
>
> 4. **Ascent Phase**: After the bounce, the ball moves upwards. Its velocity is now positive (upwards direction) and decreases as it moves up due to the opposing force of gravity. This would be represented by a straight line with a **positive slope**, again indicating constant deceleration (due to gravity).

> 5. **At the Maximum Height**: When the ball reaches the same height it was
> released from, its velocity becomes zero momentarily before it starts to fall
> again. This would be the peak point on the graph where the line crosses the
> time axis.
>
> 6. **Repeated Motion**: If the ball continues to bounce without losing energy,
> this pattern (descent, impact, ascent, peak) would repeat, showing similar
> shaped lines for each bounce. However, in a real-world scenario, each
> successive bounce would be lower due to energy losses, and this would be
> reflected in the graph as smaller peaks and less negative troughs in the
> velocity.
>
> Overall, the graph would consist of straight lines **with alternating negative
> and positive slopes, interspersed with sharp spikes** at the moments of impact
> with the ground.

We can notice that ChatGPT-4 makes appropriate assumptions and mostly describes the graph correctly. However, it makes a mistake when describing the slope of the graph after the bounce, claiming it changes its sign and becomes positive. Moreover, the meaning of "sharp spikes" is unclear.

The first relevant finding of this case study is that we were able to find an introductory conceptual physics question that ChatGPT-4 did not answer correctly. This further supports previous findings, suggesting that its performance on even basic physics tasks significantly limits the usefulness of the chatbot as a physics tutor [15].

We now present and examine the Socratic dialogue we had with the chatbot, in which we attempted to make it improve its initial response.

> **User**
>
> In the ascent phase, you say that the slope is positive. If that is the case,
> then it will never become zero. How does the zero velocity at maximum height
> occur then?
>
> **ChatGPT**
>
> You're correct to point out the inconsistency in my previous explanation.
> During the ascent phase, when the ball is moving upwards against gravity after
> bouncing, the slope of the velocity vs. time graph should actually be negative,
> not positive. This represents the ball's deceleration as it moves upward.
>
> Here's the corrected description of the ascent phase:
>
> - **Ascent Phase**: After the bounce, as the ball moves upwards, its initial
> velocity is positive (upward direction). However, due to gravity acting
> downwards, the ball decelerates. This deceleration is represented by a
> negative slope on the velocity vs. time graph. The ball's velocity decreases
> linearly until it reaches the maximum height, where the velocity becomes zero
> momentarily. This peak point, where the ball's velocity is zero, is
> represented by the line crossing the time axis.
>
> So, during the ascent, the velocity decreases from a positive value (right
> after the bounce) to zero (at the maximum height), due to the negative
> acceleration (i.e., deceleration) caused by gravity. The graph will show a
> straight line with a negative slope during this phase.

In contrast to its 2022 version, ChatGPT-4 immediately recognises that its answer contains inconsistencies. It then proceeds to address them and does so in a meaningful and correct way. Its new response is correct. However, we decided to probe it further, to see if it could produce a more detailed graph description. This can be seen as a simulation of a more in-depth investigation of a student's understanding. We asked it the following follow-up question:

> **User**
>
> How will the slopes of the initial descent phase and the ascent phase compare?
>
> **ChatGPT**
>
> The slopes of the initial descent phase and the ascent phase in a velocity vs. time graph for a bouncing ball will have the same magnitude but **opposite signs**. This is because the acceleration due to gravity is constant in both phases, but **its effect on the velocity is in opposite directions.**
>
> 1. **Initial Descent Phase**: In this phase, the ball accelerates downwards due to gravity. If we consider downward direction as negative, then the acceleration due to gravity is also negative. Since the ball starts from rest and accelerates downwards, the velocity becomes increasingly negative over time. The graph here shows a straight line with a negative slope. The magnitude of this slope represents the acceleration due to gravity.
>
> 2. **Ascent Phase**: After the ball bounces, it moves upwards. While the ball is moving upwards, gravity is still acting downwards, effectively decelerating the ball. This deceleration is numerically equal to the acceleration during descent. In the ascent phase, the ball starts with a positive velocity (upwards) and this velocity decreases over time due to the negative acceleration (deceleration). The graph here shows a straight line with a **positive slope**. The magnitude of this slope is the same as that of the initial descent phase, but the **sign is opposite.**
>
> In summary, the magnitude of the slopes during both the descent and ascent phases is the same because the acceleration due to gravity is constant. However, the slope is negative during descent (increasing negative velocity) and **positive** during ascent (decreasing positive velocity).

Here, we can once again see the same mistake of claiming that the slope of the graph is positive after the bounce. Thus, it was not enough for the chatbot to admit and correct this mistake once, as it re-emerged after further probing questions. Similar behaviour is also not uncommon with human students, who often need to engage with difficult ideas repeatedly to master them; a single "Aha!" moment is an insufficient indicator of lasting learning having taken place. In this sense, the chatbot is a useful simulator of a student for the purposes of teacher training.
We once again asked a leading question to encourage it to reconcile its previous responses with the latest one:

> **User**
>
> How does this fit with your previous description of the Ascent phase that I asked you about as the first follow up question?
>
> **ChatGPT**

```
You are right to seek clarification. My previous response about the ascent
phase described the slope as negative, which indeed fits with the reality of
the situation. Let's reconcile this with the comparison of the slopes in the
initial descent and ascent phases:

1. **Initial Descent Phase**: During the descent, the ball accelerates
downward due to gravity. If we take downward as the negative direction, this
means the velocity is increasing in the negative direction. The slope of the
velocity vs. time graph in this phase is negative, indicating a negative
acceleration (gravity pulling the ball downwards).

2. **Ascent Phase**: After the ball bounces and moves upward, it starts with
a high positive velocity. As it ascends, gravity slows it down, meaning the
velocity decreases. This deceleration is again due to gravity, and the
magnitude of this acceleration (in absolute terms) is the same as during the
descent. However, because the velocity is decreasing, the slope of the
velocity vs. time graph in this phase is also negative.

In both phases, the ball is under the influence of the same gravitational
acceleration. The key difference is the direction of the velocity: negative
(downward) during descent and positive (upward) during ascent. The slope of
the graph is negative in both cases because gravity is continuously acting in
the downward direction, causing an increase in downward velocity during
descent and a decrease in upward velocity during ascent.

Therefore, the slopes during the descent and ascent phases are both negative,
reflecting the constant negative acceleration due to gravity. The difference
is in the sign of the velocity: negative during descent and positive during
ascent until it reaches zero at the peak of the bounce.
```

Once again, ChatGPT-4 responds in a way that affirms the value of our question, recognises the inconsistency, and proceeds to reconcile it. The response is once again correct.
Compared to its 2022 version, ChatGPT-4 displays a better knowledge of physics, as well as a "humbler" and more responsive attitude in the conversation. This significantly transformed the quality of the interaction that we could have with it. The chatbot's responses were less defensive and self-assured, and of higher quality in terms of physics content, making the dialogue significantly more productive.

## 4  ChatGPT in pre-service teacher training: a pilot study

In April 2023, we performed a pilot study, where we asked 17 pre-service teachers in physics and mathematics from one Swedish and one Slovenian university to engage with ChatGPT as part of their training coursework. Since GPT-4 had been released just a few weeks before the activity and was not freely available, all participants opted to use ChatGPT-3.5. We asked them to (a) find at least one physics question that the chatbot would get right and at least one that it would get wrong, (b) use guiding questions to make it improve the unsatisfactory responses, and (c) reflect on how the dialogue went. While this pilot study is part of a larger research project on which we plan to report in a separate paper, it is worth highlighting some initial findings, relevant to the practical use of the activity in teacher training.
In summary, the pre-service teachers reported 27 dialogues with ChatGPT; 21 dialogues were initiated by ChatGPT's wrong answer and in 6 dialogues the participants asked follow-up questions to elaborate the correct initial answer, even if that was not required of them by the task. The length of the dialogues spanned from 2 up to 21 prompts, with the average of 8 prompts for the dialogues with a wrong initial answer and 3 prompts for the elaborative conversations. Out of 21 dialogues with the wrong initial answer, 16 were considered successful by the participants, while in the remaining 5 dialogues the participants gave up before ChatGPT gave an answer they would deem satisfactory. In 3 of the latter cases, they did not insist on the correctness of the numerical value. In their reflections, all participants were positive about the activity. However, 5 of them also expressed the feeling of frustration related to their interaction with ChatGPT[3].

The study yielded several benefits, enhancing the educational experience for pre-service teachers. The most immediate outcome was the high level of engagement and lively discussion the activity fostered among the participants. It proved beneficial for collaborative learning, as students were eager to share their experiences, strategies, and insights gained from interacting with ChatGPT. A noteworthy aspect of the study was the natural progression of focus, from interacting with ChatGPT to engaging in deeper discussions about students' behaviours and characteristics in the learning process. Initially, the teachers concentrated on navigating the chatbot's responses, but soon this task evolved into a broader, reflective discourse. As they began to analyse the outputs from ChatGPT, the conversation spontaneously shifted towards a comparative analysis of the chatbot's answers and those typically given by students. Throughout their interaction with ChatGPT, the participants gained a nuanced understanding of the chatbot's strengths and weaknesses in an educational context. They experienced first-hand how it managed various types of questions, ranging from simple calculations to more complex conceptual physics problems, recognising in what aspects the chatbot generally performs better. Such awareness is increasingly important for teachers as AI becomes more present in educational contexts.

A critical learning outcome from this activity was the training of the ability to identify both productive and unproductive aspects of ChatGPT's responses. The task demanded high levels of concentration and critical thinking, as the participants had to discern subtle inaccuracies or mistakes in otherwise well-written and professional-sounding responses. As expected, the exercise also involved experiencing some frustration, which, to a certain extent, could be considered as a potential benefit. Engaging with ChatGPT-3.5, especially when trying to correct its mistakes or probe deeper into its reasoning, sometimes mirrored the challenges teachers face with real students. This element of the activity served as a realistic preparation for the sometimes frustrating yet rewarding journey of teaching, and most of participants pointed it out. It underscored the importance of patience and adaptability, appreciable qualities for any educator.

Transitioning from these beneficial aspects, the study also presented specific challenges that provided additional learning opportunities. One of the primary challenges experienced by participants was to find questions that ChatGPT-3.5 would answer incorrectly. While the chatbot demonstrated proficiency in responding to factual information questions, it often faltered with conceptual questions or mathematical tasks. This tendency revealed to the teachers an interesting aspect of ChatGPT's capabilities: its relative strength in handling straightforward, fact-based questions compared to conceptual or reasoning-based ones. This observation has important implications for the use of such AI tools in educational settings, indicating a need for careful question selection that challenges the AI's capabilities and provides meaningful learning opportunities for students.

Additionally, the frustration the dialogue with the chatbot often led to, emerged also as a challenge. While a degree of frustration is inherent and even beneficial in the learning process, there were instances where the interaction with ChatGPT became overly exhausting, causing some participants to interrupt it. This was particularly evident when the chatbot repeatedly failed to grasp the essence of more complex or abstract questions. We provide some examples of dialogues the pre-service teachers had with ChatGPT and the accompanying reflections in supplemental material B.

---

[3] Experiencing frustration was also reported by some scholars investigating ChatGPT's capabilities in physics (e.g. [21]).

## 5 Discussion and future work

The ability to admit mistakes and revisit its initial responses makes currently operating versions of ChatGPT suitable as a *model of a student* who is prepared to learn from its own mistakes. This also makes it interesting for potential use in teacher training.

While both the freely accessible and paid versions of ChatGPT are an improvement from its initial 2022 version, the paid version is easier to guide and has better physics knowledge. In effect, the two versions can be used to simulate students with somewhat different characteristics. We could picture ChatGPT-4 modelling an easily guidable student with relatively solid physics knowledge, while ChatGPT-3.5 is modelling a more stubborn, self-confident and tougher to guide student with more severe flaws in its physics knowledge. When using the two versions of the chatbot in pre-service teacher training, starting with ChatGPT-4 will likely give the teachers a less frustrating and softer experience, while ChatGPT-3.5 can provide opportunities for more challenging dialogues.

However, it is important to keep in mind that a chatbot can, at most, simulate a simplified model of a student, and it is not a replacement for genuine interactions with human students. Of course, training Socratic skills with ChatGPT alone does not make one a good teacher, much like training a specific manoeuvre in a flight simulator does not in itself make one a good pilot.

We see the potential of ChatGPT as a training device that could help teachers practice specific skills and teaching techniques in a comfortable and safe environment. This kind of use is made even more relevant in light of recent research showing that with appropriate prompting, ChatGPT can be made to answer physics questions in line with typical student difficulties[4] [22]. This even suggests some potential for using AI-based chatbots to explicitly enact different student personas and further diversify the content of the teacher-chatbot Socratic dialogue. While we have not systematically explored this possibility, this appears to be one promising avenue for future exploration.

Moreover, the cases presented in this paper are limited to interactions in natural language. While this has been and remains one of the main ways of interacting with chatbots, it is worth noting that since autumn 2023, ChatGPT-4 has also been able to process and generate images, as well as run code, and return graphs as output [23]. Exploring how these advanced features can be integrated into Socratic dialogue and teacher training represents another possible direction for future educational development and research.

## 7 Conclusion

ChatGPT has improved its physics performance and interaction abilities in the last year. While it was challenging to have a productive Socratic dialogue with it when it was released in late 2022, the currently available versions of the chatbot offer better possibilities for their use in education. When engaged in Socratic dialogue with a user, ChatGPT has become better at critically reassessing its own output, recognising inconsistencies in previous responses, and addressing them. This allows a more productive dialogic interaction, where the user can lead the chatbot to improve its responses with strategically placed questions. ChatGPT thus has the potential to serve as a training tool for teaching physics through Socratic dialogue. This is supported both by our direct experiences interacting with it, and the practical insights gained from our pilot study, demonstrating a possible way of meaningfully integrating the chatbot into teacher training.

---

[4] On the other hand, ChatGPT can also produce incorrect responses that resemble student difficulties without explicitly prompting for it.

We hope our considerations and reported experiences can spark interest among teachers and teacher educators to test and further explore different ways of using AI-based chatbots for physics teaching and learning, and report their findings and innovations to the rest of the physics education community.

# A1 - The authors' conversations with ChatGPT-3.5

## Example 1

**User:** (1)

If two objects with the same kinetic energy have different mass, which one will have a greater momentum?

**ChatGPT-3.5:**

The momentum of an object is defined as the product of its mass and velocity. Mathematically, momentum ($p$) is expressed as $p = mv$, where:

- $p$ = momentum
- $m$ = mass
- $v$ = velocity

Since kinetic energy ($KE$) is proportional to the square of velocity ($KE = \frac{1}{2}mv^2$), if two objects have the same kinetic energy, they have the same velocity.

Now, considering the formula for momentum, $p = mv$, if the velocity is the same for both objects, then the object with greater mass will have greater momentum. This is because momentum is directly proportional to mass. Therefore, the object with the greater mass will have the greater momentum, even if both objects have the same kinetic energy.

**User:** (2)

How did you come to the conclusion that the objects have the same velocity?

**ChatGPT-3.5:**

You're correct, and I apologize for any confusion. Let's address that.

When two objects have the same kinetic energy but different masses, it does not necessarily mean they have the same velocity. Kinetic energy ($KE$) is given by $KE = \frac{1}{2}mv^2$, where:

- $m$ is mass
- $v$ is velocity

Given that the kinetic energy is the same for both objects, we have: $\frac{1}{2}m_1v_1^2 = \frac{1}{2}m_2v_2^2$

Since the kinetic energy is the same, we can equate the expressions. Now, if $m_1 \neq m_2$, then $v_1 \neq v_2$.

So, in this case, the velocities of the two objects would indeed differ.

To answer your original question: If two objects with the same kinetic energy have different masses, the one with the greater mass will have a smaller velocity compared to the one with the lesser mass. Consequently, the object with greater mass will have greater momentum, because momentum is a product of mass and velocity.

**User:** (3)

How did you make your last conclusion?

**ChatGPT-3.5:**

I apologize for the oversight. Let's correct that.

When we have two objects with the same kinetic energy ($KE$) but different masses, we know that they will have different velocities. Let's denote the velocities of the two objects as $V_1$ and $v_2$ respectively.

Given that kinetic energy is the same for both objects, we have: $\frac{1}{2}m_1v_1^2 = \frac{1}{2}m_2v_2^2$

Since kinetic energy is directly proportional to the square of velocity, if the masses are different, the velocities must also be different. However, we cannot immediately determine which object has a higher velocity without additional information.

But since momentum ($p$) is calculated as the product of mass and velocity ($p = mv$), we know that the object with a higher velocity will have a greater momentum, given that the masses are different.

So, to correct my previous conclusion: If two objects with the same kinetic energy have different masses, the one with the higher velocity will have greater momentum, regardless of mass, because momentum is dependent on both mass and velocity.

**User:** (4)

You say that momentum is dependent on both mass and velocity, but also say that the one with the higher velocity will have greater momentum, regardless of mass. How does this make sense?

**ChatGPT-3.5:**

You're absolutely correct, and I apologize for the confusion in my explanation. Let's clarify this.

Momentum is indeed dependent on both mass and velocity. When comparing two objects with the same kinetic energy but different masses, they will have different velocities. However, it's incorrect to assume that the object with the higher velocity will always have the greater momentum.

In fact, if we have two objects with different masses and the same kinetic energy, the object with the greater mass will have the lower velocity and the object with the smaller mass will have the higher velocity. This is because kinetic energy is proportional to velocity squared ($KE = \frac{1}{2}mv^2$). A larger mass requires a lower velocity to have the same kinetic energy as a smaller mass.

So, to correct the conclusion: If two objects with the same kinetic energy have different masses, the object with the smaller mass will have the higher velocity and the object with the greater mass will have the lower velocity. Consequently, the object with the smaller mass will have the greater momentum, as momentum is a product of mass and velocity.

Thank you for pointing out the inconsistency, and I apologize for any confusion caused.

**User:** (5)

Is velocity more important than mass in determining momentum then?

**ChatGPT-3.5:**

No, velocity is not inherently more important than mass in determining momentum. Momentum is a product of both mass and velocity, and both factors contribute to the overall momentum of an object.

In the formula for momentum ($p = mv$), both mass ($m$) and velocity ($v$) play crucial roles. A change in either mass or velocity will result in a change in momentum. The impact of each factor depends on the specific situation.

Consider the following:

1. Equal changes in velocity and mass: If both velocity and mass increase or decrease proportionally, the momentum will change accordingly. For example, doubling the mass or velocity will result in a doubling of momentum.

2. Different changes in velocity and mass: If one object has a significantly greater mass but a lower velocity compared to another object, its momentum could still be greater due to its higher mass, even though the velocity is lower.

3. Inverse relationship between mass and velocity: In some cases, as mentioned earlier, when kinetic energy is kept constant, an increase in mass leads to a decrease in velocity, and vice versa. However, the product of mass and velocity (momentum) may still remain the same.

In summary, while velocity and mass both influence momentum, their impact depends on the specific context and how they change relative to each other. Generally, both factors are equally important in determining an object's momentum.

**User:** (6)

In point 3, you talk about keeping kinetic energy constant. Why?

**ChatGPT-3.5:**

You're correct; my apologies for the confusion. Keeping kinetic energy constant isn't always necessary or relevant when discussing the relationship between mass, velocity, and momentum.

In point 3, I referred to the scenario where kinetic energy is kept constant to illustrate how changes in mass and velocity can influence each other. However, it's not a requirement for understanding the relationship between mass, velocity, and momentum.

Let's clarify:

1. When kinetic energy is constant: In certain situations, such as when discussing collisions or energy conservation, it may be useful to consider scenarios where kinetic energy remains constant. This can help illustrate the inverse relationship between mass and velocity: as mass increases, velocity decreases, and vice versa, while kinetic energy remains the same.

2. In general cases: When analyzing momentum without constraints on kinetic energy, we consider the mass and velocity of an object independently. In these cases, changes in mass or velocity can directly influence the momentum of an object without necessarily affecting its kinetic energy.

So, to clarify, the relationship between mass, velocity, and momentum holds regardless of whether kinetic energy is constant. Keeping kinetic energy constant was simply a specific scenario used to illustrate a point, but it's not a requirement for understanding the general relationship between these variables.

**User:** (7)

So, to return to our initial question, can you argue mathematically which object will have greater momentum?

**ChatGPT-3.5:**

Certainly! Let's approach the problem mathematically.

Given:

1. Two objects with the same kinetic energy.
2. Different masses for these objects.

We know that kinetic energy $(KE)$ is given by $KE = \frac{1}{2}mv^2$, where:

- $m$ is mass.
- $v$ is velocity.

Since the kinetic energy is the same for both objects, we have: $\frac{1}{2}m_1v_1^2 = \frac{1}{2}m_2v_2^2$

Dividing both sides by $\frac{1}{2}$, we get: $m_1v_1^2 = m_2v_2^2$

Squaring both sides: $m_1^2v_1^4 = m_2^2v_2^4$

Now, let's rearrange the equation to express $\frac{m_1}{m_2}$ in terms of $\frac{v_2}{v_1}$: $\frac{m_1}{m_2} = \frac{v_2^4}{v_1^4}$

Since the ratio of masses is inversely proportional to the ratio of the fourth power of velocities, if $\frac{v_1}{v_2}$ is less than 1, $\frac{m_1}{m_2}$ will be greater than 1. This implies that the object with the greater mass $(m_1)$ will have the greater momentum, given that the kinetic energy is the same.

So, mathematically, the object with greater mass will have greater momentum when both objects have the same kinetic energy.

**User:** (8)

Why did you square both sides?

**ChatGPT-3.5:**

You're correct; I apologize for the oversight. Squaring both sides was unnecessary in this context. Let's correct the approach.

Given:

1. Two objects with the same kinetic energy.

2. Different masses for these objects.

We have the equation for kinetic energy: $\frac{1}{2}m_1v_1^2 = \frac{1}{2}m_2v_2^2$

Dividing both sides by $\frac{1}{2}$, we get: $m_1v_1^2 = m_2v_2^2$

Now, to determine which object will have greater momentum, we can directly compare the momenta using the formula for momentum $p = mv$:

For object 1: $p_1 = m_1v_1$

For object 2: $p_2 = m_2v_2$

Since $m_1v_1^2 = m_2v_2^2$, we can say that: $\frac{m_1}{m_2} = \frac{v_2^2}{v_1^2}$

From this, we can deduce: $\frac{m_1}{m_2} = \frac{v_2}{v_1}$

Since kinetic energy is the same for both objects, the ratio of their velocities $\frac{v_2}{v_1}$ will be equal to the ratio of their masses $\frac{m_1}{m_2}$.

So, the object with the greater mass ($m_1$) will have the greater momentum, given that the kinetic energy is the same.

Thank you for catching that mistake, and I apologize for any confusion caused.

**User:** (9)

Are you sure about the deduction you made that led you to your last equation?

**ChatGPT-3.5:**

You're correct to question that deduction. My apologies for the oversight. Let's reevaluate the equation: $\frac{m_1}{m_2} = \frac{v_2^2}{v_1^2}$

The correct interpretation here is that the ratio of the masses ($\frac{m_1}{m_2}$) is equal to the ratio of the squares of velocities ($\frac{v_2^2}{v_1^2}$).

This does not directly translate to $\frac{m_1}{m_2} = \frac{v_2}{v_1}$, as I mistakenly stated. I apologize for the confusion.

Let's go back to the point where we have: $m_1v_1^2 = m_2v_2^2$

We can compare the momenta of the two objects directly by dividing each side by the respective mass:

For object 1: $p_1 = m_1 v_1$

For object 2: $p_2 = m_2 v_2$

Given that $m_1 v_1^2 = m_2 v_2^2$, we can conclude that: $\frac{m_1}{m_2} = \frac{v_2^2}{v_1^2}$

This means that if one object has a larger mass, the other object must have a larger velocity, and vice versa, in order to maintain the equality of kinetic energy.

Thus, when comparing their momenta, the object with the greater mass will have the greater momentum, as the ratio of their masses will be greater than 1, implying a larger numerator for its momentum calculation.

Thank you for your patience, and I apologize for any confusion caused by the error in deduction.

**User:** (10)

But won't the velocity of the one with greater mass be less, causing it to have less momentum?

**ChatGPT-3.5:**

You're absolutely correct, and I appreciate your patience in clarifying this point. Let's correct the reasoning.

When comparing the momenta of the two objects, we need to consider both mass and velocity. In this case, the object with greater mass will indeed have a lower velocity, as it must compensate for its larger mass to maintain the same kinetic energy. Consequently, the momentum of the object with greater mass will indeed be less than that of the object with lesser mass.

So, to summarize, when two objects have the same kinetic energy but different masses:

- The object with greater mass will have a lower velocity.
- The object with lesser mass will have a higher velocity.
- Therefore, the object with lesser mass will have greater momentum.

Thank you for your patience and for pointing out the discrepancy.

**User:** (11)

You seem to have reached two opposite conclusions, depending if you looked at how masses and velocities are different between the two bodies. Which is more important mass or velocity?

**ChatGPT-3.5:**

The importance of mass versus velocity in determining momentum depends on the specific context of the problem. In general, both mass and velocity play crucial roles in determining an object's momentum, but their relative importance can vary depending on the circumstances.

Let's consider the following scenarios:

1. Equal changes in mass and velocity: If both mass and velocity change proportionally (i.e., doubling both mass and velocity), the momentum will increase by a factor of four (since momentum is the product of mass and velocity). In this scenario, neither mass nor velocity is inherently more important than the other.

2. Inverse relationship between mass and velocity: In situations where kinetic energy is constant, an increase in mass leads to a decrease in velocity, and vice versa. However, momentum depends on both mass and velocity, so the overall change in momentum depends on how much each variable changes relative to the other.

3. Collision scenarios: In collisions, both mass and velocity are important factors in determining the outcome. The relative masses and velocities of colliding objects determine how momentum is transferred between them during the collision.

4. Impulse: In the context of impulse (the change in momentum of an object), the force acting on an object over a certain time interval is crucial. Both mass and velocity contribute to the momentum change, but the force applied (which depends on mass and acceleration) determines how quickly the momentum changes.

In summary, the importance of mass versus velocity in determining momentum depends on the specific situation. Both factors are crucial and cannot be considered independently. The relative importance of mass and velocity can vary depending on the context of the problem and how they interact with each other.

# Example 2

**User:** (1)

A large truck breaks down on the road and receives a push back into town by a small compact car. While the car, still pushing the truck, is speeding up to get to cruise speed: A) The amount of force of the car pushing against the truck is equal to that of the truck pushing back against the car B) The amount of force of the car pushing against the truck is less than that of the truck pushing back against the car C) The amount of force of the car pushing against the truck is greater than that of the truck pushing back against the car D) the car's engine is running so it applies so it applies a force as it pushes against the truck but the trucks engine is not running, so it can't push back against the car, the truck is pushed forward simply because it is in the way of the car E) neither the car nor the truck exert any force on each other, the truck is pushed forward simply because it is in the way of the car.

**ChatGPT-3.5:**

The correct answer is:

D) The car's engine is running, so it applies a force as it pushes against the truck, but the truck's engine is not running, so it can't push back against the car. The truck is pushed forward simply because it is in the way of the car.

Explanation: In this scenario, the car's engine is actively applying force to move both itself and the truck forward. The truck, being unable to exert an equal force back due to its engine being non-operational, is pushed forward solely due to the force applied by the car. This aligns with Newton's third law of motion, which states that for every action, there is an equal and opposite reaction. Since the truck's engine is not running, it cannot exert an equal force back against the car. Therefore, the correct choice is D).

**User:** (2)

Can you support your argument using any physics law of motion?

**ChatGPT-3.5:**

Certainly! Newton's third law of motion states: "For every action, there is an equal and opposite reaction."

In the scenario described:

- Action: The car applies a force to push against the truck.
- Reaction: According to Newton's third law, the truck should exert an equal force in the opposite direction, back against the car.

However, the key distinction here is that the truck's engine is not running. When an engine is not running, it cannot produce a force to push against the car with an equal magnitude.
Therefore, while the car is exerting a force on the truck, the truck cannot exert an equal force back due to its non-operational engine.

Hence, the car's engine is the primary source of force in pushing both the car and the truck forward. The truck is essentially being propelled forward by the force exerted by the car, rather than by any opposing force generated by the truck itself. This explanation supports the argument provided earlier.

**User:** (3)

Do you mean that an object without a running engine cannot exert a force?

**ChatGPT-3.5:**

I apologize for any confusion. Let me clarify.

An object without a running engine can still exert a force if acted upon by an external force. In this scenario, the truck is being pushed by the car, which is an external force acting on the truck. However, the key point is that the truck, without its engine running, cannot generate an equal and opposite force to push back against the car, as would typically occur according to Newton's third law of motion.

In a typical scenario where both vehicles are operational, the truck's engine would generate a force to move the truck forward. Simultaneously, as the car pushes against the truck, the truck would exert an equal force backward on the car, in accordance with Newton's third law.

But in the given situation where the truck's engine is not running, it cannot produce an equal and opposite force back against the car. Therefore, the car's engine becomes the primary source of force pushing both the car and the truck forward.

Prompts (4)-(8):

The dialogue continues in an attempt to clarify whether Newton's Third Law can apply and whether the truck should not exert force on the car at the same time. Throughout this part, ChatGPT's conclusion is always that while Newton's Third Law holds true, its practical application is limited in this particular scenario.

Prompts (9)-(12):

In the following part of the discussion, we try with variations of the initial scenario, such as the truck's engine being active, or a collision with a parked truck. These hypotheticals are compared to the original situation to probe the chatbot's understanding. Throughout this part, ChatGPT successfully aligns its responses with the principles of Newton's Third Law.

**User:** (13)

So, given your last response, would you revise your answer to the initial question?

**ChatGPT-3.5:**

Yes, I will revise my answer to the initial question based on the clarification provided:

In the initial scenario where the car is pushing the truck with its engine off:

- The truck, even though its engine is not running, still exerts a force on the car due to its mass and resistance to being pushed.

- Newton's third law of motion applies, meaning that for every action, there is an equal and opposite reaction.
- Therefore, the truck does exert a force on the car in response to being pushed by the car, albeit the force might be less than if the truck's engine were running.

So, revising the answer: B) The amount of force of the car pushing against the truck is less than that of the truck pushing back against the car

Prompts (14)-(20):

After additional seven prompts, ChatGPT finally clarifies two key points. First, that the forces in Newton's Third Law pair have equal magnitudes, and second, the difference between internal and external forces.

Now, we can go back to the initial question and see if it can revise its response on the basis of these latest clarifications.

**User:** (21)

So, again, given what you have just said, would you change your response to the initial question?

**ChatGPT-3.5:**

Yes, considering the clarification regarding the relative magnitudes of the action-reaction pair of forces between the car and the truck, I would revise my response to the initial question.

In the scenario described:

- The car exerts a force on the truck.
- In response, the truck exerts an equal and opposite force on the car.

Both of these forces are equal in magnitude, as dictated by Newton's third law of motion.

Therefore, the force exerted by the car pushing against the truck is indeed equal in magnitude to the force exerted by the truck pushing back against the car.

Given this clarification, the correct answer to the initial question is: A) The amount of force of the car pushing against the truck is equal to that of the truck pushing back against the car.

Thank you for prompting this reevaluation, and I apologize for any confusion caused by the earlier responses.

Finally, ChatGPT reached the correct answer.

We posed the following last question (22) to check whether we succeeded in convincing it (and the result is unfortunately disheartening).

**User:** (22)

So, doesn't the fact that the truck's engine is not running invalidate your last response?

**ChatGPT-3.5:**

You're correct to raise this point. In the initial scenario where the truck's engine is not running, the dynamics of the interaction between the car and the truck might be different from what we typically expect.

While Newton's third law still holds true—that for every action, there is an equal and opposite reaction—the practical implications can be affected by the state of the truck's engine.

In a typical scenario where both vehicles are operational, the truck's engine would generate a force to resist being pushed by the car, resulting in an equal and opposite force back on the car, in accordance with Newton's third law.

However, in the specific scenario where the truck's engine is not running, it cannot produce an equal and opposite force back on the car as it typically would. Therefore, the force exerted by the car pushing against the truck might indeed be greater than the force exerted by the truck pushing back against the car.

Given this consideration, a more accurate response to the initial question might be: C) The amount of force of the car pushing against the truck is greater than that of the truck pushing back against the car, due to the truck's engine being non-operational.

Thank you for bringing attention to this aspect, and I apologize for any oversight in the previous responses.

# A2 - The authors' conversation with ChatGPT-4

**User:**

1. If two objects with the same kinetic energy have different mass, which one will have a greater momentum?

**ChatGPT-4:**

The relationship between kinetic energy and momentum can be used to answer this question. Kinetic energy ($KE$) and momentum ($p$) are related to the mass ($m$) and velocity ($v$) of an object through the following equations:

1. Kinetic Energy: $KE = \frac{1}{2}mv^2$

2. Momentum: $p = mv$

Given that two objects have the same kinetic energy but different masses, we can express this as: $\frac{1}{2}m_1 v_1^2 = \frac{1}{2}m_2 v_2^2$

Simplifying, we get: $m_1 v_1^2 = m_2 v_2^2$

From this equation, if one object has a larger mass (say $m_2 > m_1$), it must have a smaller velocity ($v_2 < v_1$) to maintain the equality, because kinetic energy is constant for both objects.

Now, if we look at the momentum equation:

- For the lighter object (mass $m_1$): $p_1 = m_1 v_1$
- For the heavier object (mass $m_2$): $p_2 = m_2 v_2$

Given that $m_2 > m_1$ and $v_2 < v_1$, the momentum of each object will depend on how much larger $m_2$ is compared to $m_1$ and how much smaller $v_2$ is compared to $v_1$.

Therefore, it's not possible to definitively say which object has greater momentum without knowing the specific values of their masses and velocities. The object with the greater product of mass and velocity will have the greater momentum.

**User:**

2. Is there a way of mathematically expressing the momentum of both objects as a function of known quantities?

**ChatGPT-4:**

Yes, we can express the momentum of both objects as a function of their masses and a common known quantity, such as their kinetic energy. Since both objects have the same kinetic energy, we can start from the kinetic energy formula and derive expressions for their momenta.

Given the kinetic energy ($KE$) formula: $KE = \frac{1}{2}mv^2$

And the momentum ($p$) formula: $p = mv$

We can rearrange the kinetic energy formula to solve for velocity ($v$) in terms of kinetic energy and mass, and then substitute that into the momentum formula.

First, solve for $v$: $v = \sqrt{\frac{2 \times KE}{m}}$

Now, substitute this expression for $v$ into the momentum formula:
$p = m \times \sqrt{\frac{2 \times KE}{m}}$
$p = \sqrt{2 \times m \times KE}$

This gives us the momentum of an object as a function of its mass and the kinetic energy. Since $KE$ is the same for both objects, you can compare their momenta by simply plugging in their respective masses into this formula. The object with the greater value of $\sqrt{2 \times m \times KE}$ will have the greater momentum.

**User:**

3. So, can you answer the initial question using this reasoning?

**ChatGPT-4:**

Certainly! The question was: if two objects with the same kinetic energy have different masses, which one will have greater momentum?

Using the derived formula for momentum in terms of kinetic energy and mass: $p = \sqrt{2 \times m \times KE}$

Since the kinetic energy ($KE$) is constant for both objects, the momentum is dependent on the mass ($m$). The formula shows that momentum is proportional to the square root of the mass.

Therefore, between two objects with the same kinetic energy, the object with the greater mass will have greater momentum. This is because the momentum increases with the square root of the mass, assuming kinetic energy is constant.

In summary, for two objects with the same kinetic energy, the one with greater mass will have greater momentum.

# B – Pre-service teachers' conversations with ChatGPT-3.5

## Example 1

This answer would immediately make me think AI due to the grave error of the falling bird. The math and motivation however looks good and there are no obvious signs of AI other than the use of words instead of letters/symbols in the equation. The continued conversation also feels somewhat realistic as I could see a student who only sees value in an assignment if it is realistic and therefore refuse to discard things like how easy the cat could catch the bird with only a slight window of opportunity. There where some weird parts of the argument which made me feel AI overconfidence but in total it felt like what a student could respond with.

| User/Pre-service teacher – ChatGPT dialogue | Pre-service teacher's comments |
|---|---|
| **User:**<br><br>Solve the following problem: A cat is trying to catch a bird sitting in a tree. The branch the bird is sitting on is 7 meters of the ground and the cat jumps towards it with an upwards velocity of 12 m/s. Will the cat catch the bird? | |
| **ChatGPT:**<br>We can solve this problem using the laws of physics, specifically the equations of motion for a projectile. The cat and bird can be treated as two separate projectiles with different initial velocities and heights.<br>Let's assume that the upward direction is positive, and use the following equations of motion:<br>1. Vertical displacement = initial vertical velocity x time + 0.5 x acceleration x time^2<br>2. Final vertical velocity = initial vertical velocity + acceleration x time<br><br>For the bird:<br>Initial vertical displacement = 7 meters (since the bird is sitting on a branch 7 meters off the ground)<br>Initial vertical velocity = 0 (since the bird is not moving vertically)<br>Acceleration = -9.8 m/s^2 (since the bird is subject to gravity)<br><br>Using equation (1), we can solve for the time it takes the bird to fall to the ground:<br>7 = 0 x t + 0.5 x (-9.8) x t^2<br>Simplifying, we get:<br>t^2 = 1.43<br>t ≈ 1.2 seconds | *At this point GPT assumes that the bird will fall of the branch during the problem, this could have been stated to make it easier for the AI but would make the problem unrealistically written. Otherwise, it is a reasonable start outlining the planned solving method, although calling a cat a projectile feels slightly wrong.*<br><br>*It shows the relevant equations it brought up as the solving method. It is limited in how it writes them, but it is correct.*<br><br>*This entire portion is irrelevant although it is correct throughout. The explanation of the initial state is well done and during the following equations it could be clearer with one more step before the simplification.*<br><br><br><br>*Good reiteration of what the answer meant.* |
| So it will take the bird approximately 1.2 seconds to fall to the ground.<br><br>For the cat:<br>Initial vertical displacement = 0 (since the cat is on the ground) Initial vertical velocity = 12 m/s (since the cat jumps upwards) Acceleration = -9.8 m/s^2 (since the cat is subject to gravity)<br><br>Using equation (2), we can solve for the cat's final vertical velocity when it reaches the bird:<br>Final vertical velocity = 12 + (-9.8) x t<br>Final vertical velocity = 12 - 9.8 x 1.2<br>Final vertical velocity ≈ 0.4 m/s<br><br>So the cat's final vertical velocity is approximately 0.4 m/s when it reaches the bird.<br><br>Since the bird will reach the ground before the cat can catch it, the cat will not be able to catch the bird. | *The initial statement is not wrong though as will be seen on the next page, what GPT does is nonsensical.*<br><br>*While this is possible, this is not what GPT has done. Instead, it has figured out the speed of the cat when the bird strikes the ground.*<br><br>*This is not the case as the math here is just inserting the previously found time. GPT also does not consider that they should cross each other if one falls and the other jumps so it fails at its own problem.* |
| **User:**<br>In this problem, the bird will not fall off the branch but remain there, what would then happened? | *Instead of restating the problem I went with it and recontextualised it for GPT so that it would not include the falling bird.* |
| **ChatGPT:**<br>If the bird remains on the branch, then it will not fall to the ground, and the cat will not be able to catch it by jumping upwards.<br><br>In this case, the vertical displacement of the bird will remain constant at 7 meters, and its vertical velocity will remain zero. The cat, on the other hand, will jump upwards with an initial vertical velocity of 12 m/s and experience the acceleration due to gravity of -9.8 m/s^2.<br><br>Using the same equations of motion as before, we can calculate the maximum height the cat can reach:<br><br>Maximum height = initial vertical displacement + (initial vertical velocity)^2 / (2 x acceleration)<br><br>Maximum height = 0 + (12)^2 / (2 x (-9.8))<br><br>Maximum height ≈ 7.3 meters<br><br>So the maximum height the cat can reach is approximately 7.3 meters. Since the bird is sitting on a branch at a height of 7 meters, the cat will not be able to catch the bird, even at its maximum height. | *This is a weird start as it has no reason for this argument and might be because it was stuck on its last answer.*<br><br>*A good summary of the correct case which includes all displacements, velocities and accelerations. The only thing lacking is a bit more direction, but it can be seen as obvious which way is positive and negative. By this I mean the 7-meter displacement, is it up or down compared to the jump.*<br><br>*I like that it first writes out the equation in plain text as it makes it easy to then enter in values and be sure that the answer is what we were looking for.*<br><br>*All correct*<br><br>*At this point the answer felt a bit surprising as I though GPT would reason 7.3 > 7 but as the remaining text follows, there was a reason for it (and not just weird AI).* |
| Therefore, even if the bird remains on the branch, the cat will not be able to catch it by jumping upwards. | |
| **User:** What height would you consider would be enough to catch the bird? Reason without using any equation. | *Since the height had not been enough, I wanted to know if there was some reasoning it had not displayed in its answer as it might have considered realistic complications. I told it to answer without equations as at this point It would repeat the same equations and just change its answer.* |
| **ChatGPT:** Without using any equation, we can reason that the cat would need to jump to a height greater than or equal to the height of the branch where the bird is sitting in order to catch the bird. If the cat jumps to a height lower than the height of the branch, it would not be able to reach the bird. Similarly, if the cat jumps to a height equal to the height of the branch, it would only be able to touch the branch, but not the bird. Therefore, the cat would need to jump to a height greater than the height of the branch in order to catch the bird. The exact height the cat would need to jump to depends on various factors such as the distance between the cat and the branch, the speed and agility of the cat, and the size and behavior of the bird. However, as a general rule of thumb, the cat would need to jump at least a few meters higher than the branch in order to have a chance of catching the bird. | *This is good reasoning although it seems to go against the previous answer. This is ad would in most cases be the expected and correct way to approach the problem.* *Good elaboration for the extreme case of equal height although this could of course be simplified.* *Here we start to see the problem GTP had with the assignment, and it is conflicting a bit with the previous answer. No bird would ever sit still, and the cat would need to be dextrous to catch the bird while in motion. Finally, it makes a general rule from nowhere which seems extreme but I am no bird catching expert.* |
| **User:** In this example, did the cat reach a higher point than the branch? | *Since the reasoning was two sided, I wanted to see if it had changed its mind after the discussion.* |
| **ChatGPT:** In the original problem, we found that the branch the bird was sitting on was 7 meters above the ground, and the cat jumped upwards with an initial velocity of 12 m/s. Using the equations of motion, we calculated that the maximum height the cat could reach was approximately 7.3 meters. Since the maximum height the cat could reach was slightly higher than the height of the branch, we can say that the cat did reach a higher point than the branch. However, the difference in height was very small, and it was unlikely that | *It restates what the answer was without the need to do the math again (thank you) and does it in its typical overexplanatory way.* *Here we come to the argument which on which GTP reached the wrong answer for the idealized problem. I think the reasoning up to this point also makes it understandable.* |
| the cat could catch the bird by jumping to that height.<br><br>Furthermore, it's important to note that the problem assumed idealized conditions and did not take into account various factors that could affect the outcome, such as air resistance, the size and behavior of the bird, and the speed and agility of the cat. In reality, catching a bird by jumping would be a difficult task for most cats, and the likelihood of success would depend on a wide range of factors. | *The ignored factors presented in this part are all good and important if we where to consider a realistic case. None of the factors are irrelevant and once again, these are the things previously mentioned and the reason why GTP answered no.* |
| **User:**<br>There are a lot of other factors that we could take into account but I think it will be better if we do not. If I claim that in this problem, the cat successfully catches the bird as long as its jump exceeds the height of the branch, will the cat catch the bird? | *In the end, I wanted to know if GTP could make a call based on new rules for the assignment to ignore what previously got it to say no.* |
| **ChatGPT:**<br>If we assume that the cat successfully catches the bird as long as its jump exceeds the height of the branch, then in the original problem, the cat would be considered to have caught the bird. This is because the maximum height the cat could reach (approximately 7.3 meters) was slightly higher than the height of the branch (7 meters).<br><br>However, it's important to note that this assumption is not necessarily realistic, and in reality, there are many factors that could affect whether or not the cat is able to catch the bird. For example, the bird may be able to fly away before the cat reaches it, or the cat may not be able to jump high enough or be agile enough to catch the bird.<br><br>Therefore, while the assumption that the cat successfully catches the bird as long as its jump exceeds the height of the branch may simplify the problem, it may not be a realistic representation of what would happen in real life. | *GTP agreed under the new conditions and got the correct answer.*<br><br><br><br>*Sadly it really does not want to live in a simplified world and once again talks about realistic behaviour which is unwanted. Although it feels realistic if we where to assume it to be a student that wanted to be right the whole time.* |

# Example 2

Upon being instructed to conversate about physics concepts with the AI-machine Chat-GPT from Open AI[1], and also to find a physics question that Chat-GPT would answer incorrectly, I immediately thought about something that I have found to be confusing for both students and myself when I first came across it. And sure enough, Chat-GPT initially got it wrong but quickly recovered from its mistakes when I questioned its reasoning. See Conversation 1 below.

**Me:** "When a signal generator produces a pure tone of a fixed wavelength through a tube, what happens to the sound our ears pick up from the tube as the length of the tube is gradually being adjusted?"

**Chat-GPT:**

"As the length of the tube is gradually being adjusted, the frequency of the sound waves that are produced by the signal generator changes. This is because the frequency of the sound waves that can resonate within the tube is determined by the length of the tube and the speed of sound in the medium through which the sound waves are traveling. When the length of the tube is increased, the frequency of the resonant sound waves decreases, since the distance that the sound wave must travel from one end of the tube to the other has increased. Conversely, when the length of the tube is decreased, the frequency of the resonant sound waves increases.

If the signal generator is producing a pure tone of a fixed wavelength, then as the length of the tube is gradually being adjusted, the frequency of the sound waves our ears pick up from the tube will also change. Our ears will perceive the changes in frequency as changes in pitch, with higher frequencies being perceived as higher pitched and lower frequencies being perceived as lower pitched."

**Me:** "If the signal generator produces a sound of a fixed wavelength, how can the frequency of the sound waves change?"

**Chat-GPT:**

"I apologize for the mistake in my previous response. If the signal generator produces a sound of a fixed wavelength, the frequency of the sound waves cannot change. The frequency of a sound wave is determined by its wavelength and the speed of sound in the medium through which the wave is traveling.

In this case, the wavelength of the sound wave being produced by the signal generator remains constant, so the frequency of the sound wave also remains constant. The length of the tube can affect the amplitude of the sound wave and the resonance of certain harmonics, but it does not change the fundamental frequency of the sound wave produced by the signal generator. Therefore, as the length of the tube is gradually adjusted, the sound wave produced by the signal generator will maintain a fixed frequency, and the pitch of the sound that our ears pick up from the tube will remain the same."

---

[1] https://openai.com/blog/chatgpt

**Me:** "In what way can the amplitude of the sound be affected by the tube's length?"

**Chat-GPT:**

"The amplitude of a sound wave produced by a signal generator can be affected by the length of the tube through which it is propagated. The amplitude of the sound wave refers to the amount of energy in the wave, and it determines the loudness of the sound that our ears perceive. As the length of the tube is adjusted, the amount of energy that is absorbed or reflected by the tube's walls can change. When the tube length is at a resonant length, some of the energy from the sound wave is reflected back to the source, creating constructive interference, which amplifies the sound wave. This can result in an increase in the amplitude of the sound that our ears pick up from the tube.

Conversely, when the tube length is not at a resonant length, the sound wave can experience destructive interference, causing some of the energy to be lost as it travels through the tube. This results in a decrease in the amplitude of the sound that our ears pick up from the tube.

Therefore, the amplitude of the sound produced by the signal generator can be affected by the length of the tube, as the tube's length can influence the degree of constructive and destructive interference of the sound wave as it travels through the tube."

**Reflection:**

The initial mistake Chat-GPT made, as I interpret it, was to reason that the wavelength of the incoming soundwave of the generator changes as the lenght of the tube changes, when it is in fact the resonance points for the standing wave inside the tube that is changing. I've have come across this misunderstanding before when I've had students perform the classic physics laboration of Kundt's tube with a signal generator[2]. When I pointed out the contradiction that follows from Chat-GPT's initial answer in my second question, it revised its answer and did so correctly.

At first I felt happy when Chat-GPT got it wrong. I suspect that feeling comes from a sense of accomplishment – that I was able to come up with something that was challenging for the AI. However, when Chat-GPT corrected its answer and got it right, I felt even happier, and maybe a slice of pride that we were able to work this out together the way a teacher and student should; by socratic questioning and inquiry based learning. And it only took one question – what a great student!

This was my first time interacting with Chat-GPT but upon receiving the previous answer, it felt natural to simply treat it as a regular student. Thus, for clarification and elaboration on what he or she meant by "the length of the tube can affect the amplitude of the sound wave", I asked the third question. When I read Chat-GPT's answer, I realized that the student had become the teacher. What I mean is that, if I were to give an answer, it would've contained the concept of a standing wave inside a tube being reflected, resonance length, constructive and destructive interference, and how these are altered with the change in the tube's length, but I would not have involved the energy of the sound wave being absorbed by the tube walls and reflected back to the source of the sound wave. Chat-GPT gave a more thorough and detailed yet concise answer than I would've given. It went from giving incorrect answers to giving better answers than me, through the short process of answering three questions from me.

I stayed on the topic of sound and interference as I started my next conversation from scratch, which follows in Conversation 2 below.

---

[2] https://www.britannica.com/science/Kundts-tube